\documentclass[useAMS,usenatbib,usegraphicx,twocolumn]{mn2e}
\def\be{\begin{eqnarray}}
\def\ee{\end{eqnarray}}

\def\h{_{\rm h}}

\def\bh{_{\rm BH}}
\def\sph{_{\rm sph}}
\def\nsph{_{\rm nsph}}
\def\calJ{{\cal J}}

\def\kms{{\rm\,km\,s^{-1}}}

\def\msun{{\rm\,M_\odot}}

\def\kpc{{\rm\,kpc}}
\def\yr{{\rm\,yr}}
\def\Myr{{\rm\,Myr}}
\def\pc{{\rm\,pc}}

\def\au{{\rm\,AU}}

\def\bulge{_{\rm bulge}}
\def\disk{_{\rm disk}}
\def\sub{{_{\rm sub}}}
\def\halo{_{\rm halo}}

\title[Kinematics of hypervelocity stars]{Kinematics of hypervelocity stars in the triaxial halo of the Milky Way}
\author[Q.\ Yu \& P.\ Madau]{Qingjuan Yu\thanks{Also a Hubble Fellow at the Department of Astronomy, University of California at Berkeley, Berkeley, CA 94720.}
and Piero Madau\thanks{Also at the Max-Planck-Institut f\"ur Astrophysik,
Karl-Schwarzschild-Str. 1, 85740 Garching, Germany.}\\
Department of Astronomy and Astrophysics, University of California, Santa Cruz, CA 95064\\
yqj@ucolick.org; pmadau@ucolick.org}

\begin{document}

\label{firstpage}
\maketitle

\begin{abstract}
Hypervelocity stars (HVSs) ejected by the massive black hole at the 
Galactic center have unique kinematic properties compared to other halo stars. Their 
trajectories will deviate from being exactly radial because of the asymmetry 
of the Milky Way potential produced by the flattened disk and the triaxial dark 
matter halo, causing a change of angular momentum that can be much larger than 
the initial small value at injection. We study the kinematics of HVSs and
propose an estimator of dark halo triaxiality that is determined only by 
instantaneous position and velocity vectors of HVSs at large Galactocentric
distances ($r\ga 50\,\kpc$). We show that, in the case of a substantially triaxial halo, 
the distribution of deflection angles (the angle between the stellar 
position and velocity vector) for HVSs on bound orbits is spread uniformly 
over the range 10$^\circ$--180$^\circ$. Future astrometric and deep wide-field 
surveys should measure the positions and velocities of a significant number of 
HVSs, and provide useful constraints on the shape of the Galactic dark matter
halo.
\end{abstract}
\begin{keywords}
black hole physics -- Galaxy: center -- Galaxy: halo -- stellar dynamics
\end{keywords}

\section{Introduction}\label{sec:intro}

Recent observations have revealed the existence of a population of 
hypervelocity stars (HVSs) traveling in the halo of the Milky Way (MW) with 
Galactic rest-frame velocities $v_{\rm rf}$ in the range between $+400$ and $+750\,\kms$ 
\citep{Brown05,Edelmann05,Hir05,Brown06a,Brown06b}. HVSs are probably B-type 
main sequence stars with lifetimes $\la 100\,\Myr$, Galactocentric distances 
$>50$ kpc, and move with speeds large enough to escape from the Galaxy. 
The significant excess of B-type stars with velocities $+275<v_{\rm 
rf}<+450\,\kms$ and distances $>10$ kpc observed by \citet{Brown07} may also
be an indication that many HVSs are ejected into the halo on {\it bound} orbits.  

HVSs were first recognized by \citet{H88} as an unavoidable byproduct of the presence 
a massive black hole (BH) in the Galactic center. Only a close encounter with 
a relativistic potential well can accelerate a 3-4 $\msun$ star to such extreme 
velocities, and at least three different ejection mechanisms have been 
proposed: the interaction between background stars and an intermediate-mass black 
hole (IMBH) inspiralling towards Sgr A$^*$ \citep{YT03,Levin06,BGPZ06,SHM06}, the 
disruption of stellar binaries in the tidal field of Sgr A$^*$ 
\citep{H88,YT03,GL06,Bromley06}, and the scattering of stars off a cluster 
of stellar-mass BHs orbiting Sgr A$^*$ \citep{OL06}
In all these models, 
HVSs have unique kinematics compared to other halo stars: 1) they have almost zero 
initial specific angular momentum at ejection, $\sqrt{GM\bh r_p}
\simeq 4.0\times 10^{-6}\kpc^2\Myr^{-1} (M\bh/3.6\times 10^6\msun)^{1/2}
(r_p/10^{-6}\kpc)^{1/2}$, where $M\bh$ the mass of Sgr $A^*$ and $r_p$ 
the pericenter distance of the star;
2) their high speeds diminish the impact of two-body relaxation or dynamical 
friction effects on their motion; and 3) their trajectories will deviate from 
being exactly radial because of the asymmetry of the Milky Way potential 
produced by the flattened disk and the triaxial dark matter (DM) halo, causing a 
change of angular momentum that can be much larger than the initial small value.
(For reference, a $1\,\kms$ deviation of the velocity from the radial
direction at $50\kpc$ represents a change of $0.05\kpc^2\Myr^{-1}$ in specific
angular momentum.) Proper-motion measurements of HVSs may therefore become
a key diagnostic tool for constraining the shape of the Galactic 
potential \citep{Gnedin05}. 

Triaxial halos are a generic prediction of the hierarchical, cold 
dark matter (CDM) models of structure formation. Dissipationless cosmological 
simulations typically predict minor-to-major density axis ratios in the range 
0.4-0.8 (e.g. \citealt{JS02}), with the asphericity of the potential increasing
rapidly towards the center of the halo \citep{HNS06}. Gas cooling tends
to circularize the potential (e.g. \citealt{Dub94,Kazant04}), while subsequent 
mergers produce highly elongated remnants (e.g. \citealt{Moore04}).
Studies of weak gravitational lensing and X-ray observations of elliptical 
galaxies show that halos are significantly flattened, in fair agreement with results 
from numerical simulations \citep{Hoe04,Buo02}. Yet the coherence of tidal debris
from the Sagittarius dwarf galaxy appears to indicate that the inner halo of
the MW is nearly spherical and therefore in conflict with CDM predictions 
(Ibata et al. 2001; but see Helmi 2004). 

In this paper, we study the kinematics of HVSs in the MW as a probe of the 
triaxiality of the Galactic halo. The outline is as follows. In 
\S~\ref{sec:analysis}, we analyze the motion of HVSs in a flattened or 
triaxial gravitational potential. We provide a
concise statistical estimator for the triaxiality of the Galactic halo
potential through the measured angular momenta of HVSs.  In \S~\ref{sec:potential}, we
review the Galactic potential model to be used in our calculations. In
\S~\ref{sec:simulation} we perform numerical simulations of the motion of HVSs to
study their kinematics. Finally, in \S~\ref{sec:conclusion}, we summarize our 
conclusions.

\section{Motion of hypervelocity stars}\label{sec:analysis}

Consider a star with position vector $\vec{r}$ moving with velocity $\vec{v}$ in 
a gravitational potential 
$\Phi=\Phi\sph(r)+\Phi\nsph(x,y,z)$, where $\Phi\sph$ and $\Phi\nsph$ are 
the spherically-symmetric and aspherical component of the the potential, 
$(x,y,z)$ are Cartesian coordinates, and $r=\sqrt{x^2+y^2+z^2}$. The rate of 
change of the specific angular momentum of the star, $\vec{J}=\vec{r}\times \vec{v}$,
is equal to the torque, 
\be
d\vec{J}/dt=-\vec{r}\times\nabla\Phi=-\vec{r}\times\nabla\Phi\nsph,
\label{eq:dJdt}
\ee
and has components
\be
\cases{
dJ_x/dt=2yz\left(\partial\Phi\nsph/\partial y^2-\partial\Phi\nsph/\partial z^2\right), \cr
dJ_y/dt=2xz\left(\partial\Phi\nsph/\partial z^2-\partial\Phi\nsph/\partial x^2\right), \cr
dJ_z/dt=2xy\left(\partial\Phi\nsph/\partial x^2-\partial\Phi\nsph/\partial y^2\right).
}
\label{eq:dJidt}
\ee
It is convenient to change from Cartesian to spherical coordinates,
$(x,y,z)=(r\sin\theta\cos\phi,r\sin\theta\sin\phi,r\cos\theta)$,
and combine the above equations to yield 
\be
\frac{dJ_x/dt}{\sin\theta\sin\phi\cos\theta}+\frac{dJ_y/dt}{\sin\theta\cos\phi\cos\theta}+\frac{dJ_z/dt}{\sin^2\theta\cos\phi\sin\phi}=0.
\label{eq:dJ0}
\ee
From the definition of angular momentum it is also easy to derive
\be
\calJ_x+\calJ_y+\calJ_z=0,
\label{eq:calJ0}
\ee
where 
\be
\cases{
\calJ_x\equiv J_x/(\sin\theta\sin\phi\cos\theta),\cr
\calJ_y\equiv J_y/(\sin\theta\cos\phi\cos\theta),\cr
\calJ_z\equiv J_z/(\sin^2\theta\cos\phi\sin\phi) 
\label{eq:calJ}
}
\ee
are determined directly from the position and velocity of the star. Note 
that equations (\ref{eq:dJ0}) and
(\ref{eq:calJ0}) are rotationally invariant, that is, they do not change when
arbitrary rotations are applied to their arguments. Below we apply the above analysis to 
the motion of stars in two simple cases of non-spherical potentials.
\begin{itemize}
\item If the non-spherical component of the gravitational potential is
axisymmetric about the plane $z=0$,
\be
\Phi\nsph=\Phi\nsph(R=\sqrt{x^2+y^2},z),
\label{eq:Phiaxis}
\ee
then $\partial\Phi\nsph(R,z)/\partial x^2=\partial\Phi\nsph(R,z)/\partial y^2$, and $J_z$
is conserved. Stars ejected from the Galactic center on radial orbits move in a plane
with 
\be
\calJ_x=-\calJ_y, \qquad \calJ_z=0.
\label{eq:calJaxisy}
\ee
\item If the non-spherical component of the potential is
triaxial, 
\be
\Phi\nsph=\Phi\nsph(x^2+y^2/p^2+z^2/q^2),
\ee
then a triaxiality parameter can be defined as
\be
T\equiv {p^{-2}-1 \over q^{-2}-1}.
\label{eq:Tdef}
\ee
If $p=q=1$, the potential reduces to the spherical case. If $p=1$ and $q\ne1$ ($T=0$),
$q=1$ and $p\ne1$, or $p=q\ne1$ ($T=1$), the potential is axisymmetric. 
If $q<p<1$, the triaxiality parameter is $0<T<1$. In a triaxial potential,
equation (\ref{eq:dJidt}) can be written as 
\be
\frac{dJ_z/dt}{\sin^2\theta\cos\phi\sin\phi}=-T \frac{dJ_y/dt}{\sin\theta\cos\phi\cos\theta}.
\label{eq:TdJ}
\ee
For HVSs moving away from the Galactic center on radial orbits,
the deviation of their trajectory from the initial ejection direction,
($\delta\theta,\delta\phi$), is small. Replacing the angles $(\theta,\phi)$ in 
equation (\ref{eq:TdJ}) with $(\theta\pm\delta\theta,\phi\pm\delta\phi)$ and integrating
yields
\be
T &= &-\frac{\calJ_z}{\calJ_y}[1\mp\frac{2\delta\theta}{\sin(2\theta)}\mp\frac{\delta\phi}{\tan\phi}+\frac{\delta^2\theta}{\sin^2\theta}+\frac{\delta^2\phi}{\sin^2\phi}+ \nonumber \\
  & & \frac{2\delta\theta\delta\phi}{\sin(2\theta)\tan\phi}+...],
\label{eq:TcalJ}
\ee
where the $\sin$ and $\cos$ arguments have been kept constant in the integration.
The term in parenthesis specifies the maximum systematic error on the triaxiality 
estimator $T=-\calJ_z/\calJ_y$, and numerical calculations (see \S\ 4 below) of the
motion of HVSs in a triaxial potential show that the typical error on $T$ is actually
smaller. For a sample of $N$ 
HVSs we can use equation (\ref{eq:TcalJ}) to construct a statistical estimator 
of triaxiality as 
\be
\bar{T}=\sum_{i=1}^{N}\frac{T_i}{\sigma^2_{T_i}}/\sum_{i=1}^{N}\frac{1}{\sigma^2_{T_i}},
\label{eq:Tbar}
\ee
with standard deviation
\be
\sigma_{\bar{T}}=\left(\sum_{i=1}^{N}\frac{1}{\sigma^2_{T_i}}\right)^{-1/2}.
\label{eq:Terror}
\ee
Here 
\be
T_i\equiv-\frac{\calJ_{z,i}}{\calJ_{y,i}}=-\frac{J_{z,i}\cos\theta_i}{J_{y,i}\sin\theta_i\sin\phi_i},
\ee
and $\sigma_{T_i}$ is its error. Note that the parameters $T_i$ are fully determined by the 
instantaneous positions and velocity vectors of the HVSs in the sample, and that, while
the ratio $-\calJ_{z,i}/\calJ_{x,i}$ can also be used to estimate the triaxiality 
$(1-p^{-2})/(q^{-2}-p^{-2})$, this does not provide any independent 
information since it can be trivially derived from $\calJ_{z,i}/\calJ_{y,i}$ 
using equation (\ref{eq:calJ0}).

In the following we set the $z$-axis of the triaxial potential to be normal
to the Galactic disk plane, and denote with $\eta_0$ the angle measured 
counter-clockwise from a reference direction (e.g. the line from the Galactic
center to the Sun) to the $x$-axis of the potential. The ratio $-\calJ_{z,i}'/\calJ_{y,i}'$ 
in a frame forming an angle $\eta$ with the reference direction can be written as 
\be
-\frac{\calJ_{z,i}'}{\calJ_{y,i}'}=\frac{\frac{1}{p^2}-1}{\left(\frac{1}{q^2}-1\right)A-\left(\frac{1}{p^2}-1\right)B},
\ee
where
\be
A=\frac{\sin(2\phi_i')}{\sin[2(\eta-\eta_0+\phi_i')]},\qquad
B=\frac{\sin(\eta-\eta_0)\sin\phi_i'}{\cos(\eta-\eta_0+\phi_i')}.
\ee
If $\eta\ne\eta_0$, the values $-\calJ_{z,i}'/\calJ_{y,i}'$ may spread out over a large
range due to the different angles $\phi_i'$ in the sample. The angle $\eta_0$ can
be estimated by minimizing the weighted variance of
$T_i'\equiv -\calJ_{z,i}'/\calJ_{y,i}'$:  
\be
\sum_{i=1}^{N}\left(\frac{T_i'-\bar{T'}}{\sigma^2_{T_i'}/\sigma^2_{\bar{T}'}}\right)^2,
\label{eq:variance}
\ee
where $\sigma_{T_i'}$ is the error of $T_i'$ and the values of $\bar{T'}$ and
$\sigma^2_{\bar{T}'}$ are obtained from equations (\ref{eq:Tbar}) and
(\ref{eq:Terror}).
\end{itemize}

\section{Galactic gravitational potential}\label{sec:potential}

We use here a four-component model for the gravitational potential of the Milky Way,
$\Phi=\Phi\bh+\Phi\bulge+\Phi\disk+\Phi\halo$, where (cf. \citealt{Gnedin05}):
\begin{itemize}

\item $\Phi\bh$ is the contribution of Sgr A$^*$, 
\be
\Phi\bh=-\frac{GM\bh}{r},
\label{eq:Phibh}
\ee
with mass $M\bh\simeq 3.6\times 10^6\msun$ \citep{Ghez05,Eisenhauer05}.
The radius of influence of Sgr A$^*$ is $GM\bh/\sigma_c^2\simeq 1.6\pc\ 
(M\bh/3.6\times10^6\msun) (100\kms/\sigma_c)^2$,
where $\sigma_c$ is the one-dimensional stellar velocity dispersion in the
Galactic center.

\item $\Phi\bulge$ is the contribution of the spherical bulge \citep{Hern90},
\be
\Phi\bulge=-\frac{GM\bulge}{r+a\bulge},
\label{eq:Phib}
\ee
with mass $M\bulge=10^{10}\msun$ and core radius $a\bulge=0.6\kpc$. As both 
the bulge mass and size are small compared to those of the
Galactic disk and halo, a slight deviation of the bulge from 
sphericity will not have a significant effect on the change of angular
momentum of HVSs in the halo. 

\item $\Phi\disk$ is the contribution of the axisymmetric disk \citep{MN75},
\be
\Phi\disk(R,z)=-\frac{GM\disk}{\sqrt{R^2+\left(a\disk+\sqrt{z^2+b\disk^2}\right)^2}},
\label{eq:Phid}
\ee
with mass $M\disk=4\times 10^{10}\msun$, scale length $a\disk=5\kpc$, and 
scale height $b\disk=0.3\kpc$.

\item $\Phi\halo$ is the contribution of the triaxial dark matter halo,
\begin{eqnarray}
\Phi\halo(x,y,z)& = &\Phi_{\rm NFW}(r^t),\\
r^t& = &p^{1/3}q^{1/3}\left(x^2+\frac{y^2}{p^2}+\frac{z^2}{q^2}\right)^{1/2}, \label{eq:Phihpq}\\
\Phi_{\rm NFW}(r^t)& = &-\frac{GM_{200}}{r_sf(c)}\frac{\ln(1+r^t/r_s)}{r^t/r_s}. 
\end{eqnarray}
Here $c\equiv r_{200}/r_s$ is the halo concentration parameter, $r_{200}$ the radius 
within which the enclosed average density is 200 times the mean matter density, 
$r_s$ the scale radius, and $f(c)=\ln(1+c)-c/(1+c)$. This generalization of an NFW 
\citep{NFW96} model ensures that
the spherically-averaged potential of the triaxial halo is similar to that of
a spherical halo with the same mass $M_{200}$ and scale radius \citep{HNS06}.
We choose the following parameters for the Milky Way halo: $c=15$,
$r_{200}=389\kpc$, $M_{200}=1.8\times10^{12}\msun$ (e.g. \citealt{Diemand07}), and assume
that the $z$-axis of the halo and disk potentials coincide.
We also assume, for simplicity, that the asphericity of the potential is
constant with radius $r$, and set $p=0.8$ and $q=0.7$ in the calculations below.

\end{itemize}

CDM halos are not smooth but have a wealth of substructure on all resolved mass
scales (e.g. \citealt{Moore99,Klypin99}). The assumption made above of a 
smooth gravitational potential neglects the deflection of HVS trajectories by halo 
substructure.  A star with velocity $v$ passing within a distance
$r_* $ from a subhalo of mass $M\sub$ will change its velocity
by an amount $\delta v=2GM\sub/(vr_*)$. The mass function of substructure in
a Milky Way-sized halo can be described as $N(>M\sub) =6.4\times 10^{-3}
(M_{200}/M\sub)$ in the subhalo mass range
$10^6\msun<M\sub<f_{\rm max}M_{200}$ \citep{Diemand07}, with $f_{\rm max}=0.01$. 
The probability that a HVS ejected from the Galactic center is subject to a 
velocity deflection $>\delta v$ can then be estimated as 
\begin{eqnarray}
P(>\delta v)&=&{3\over 4}\int \frac{r^2_*}{r_{200}^2}\,\frac{dN}{dM\sub}
dM\sub \nonumber \\
&\sim& 0.08 \left(\frac{f_{\rm max}}{0.01}\right)\,\left(\frac{V_c}{140\,\kms}\right)^4
\times \nonumber \\
& & \left(\frac{10^3\kms}{v}\right)^2\left(\frac{1\kms}{\delta v}\right)^2,
\end{eqnarray}
where $V_c\equiv (GM_{200}/r_{200})^{1/2}$ is the halo circular velocity. This
probability is quite low so that a smooth potential is a good assumption in
this work.

\section{Numerical calculations}\label{sec:simulation}

In this section we perform numerical calculations of the motion of HVSs
in the MW gravitational potential. 

\begin{figure*}
\begin{center}
%\epsscale{0.8} \plotone{map.epsi}
%\includegraphics[width=0.85\textwidth,angle=0]{map.epsi}
\includegraphics[width=0.85\textwidth,angle=0]{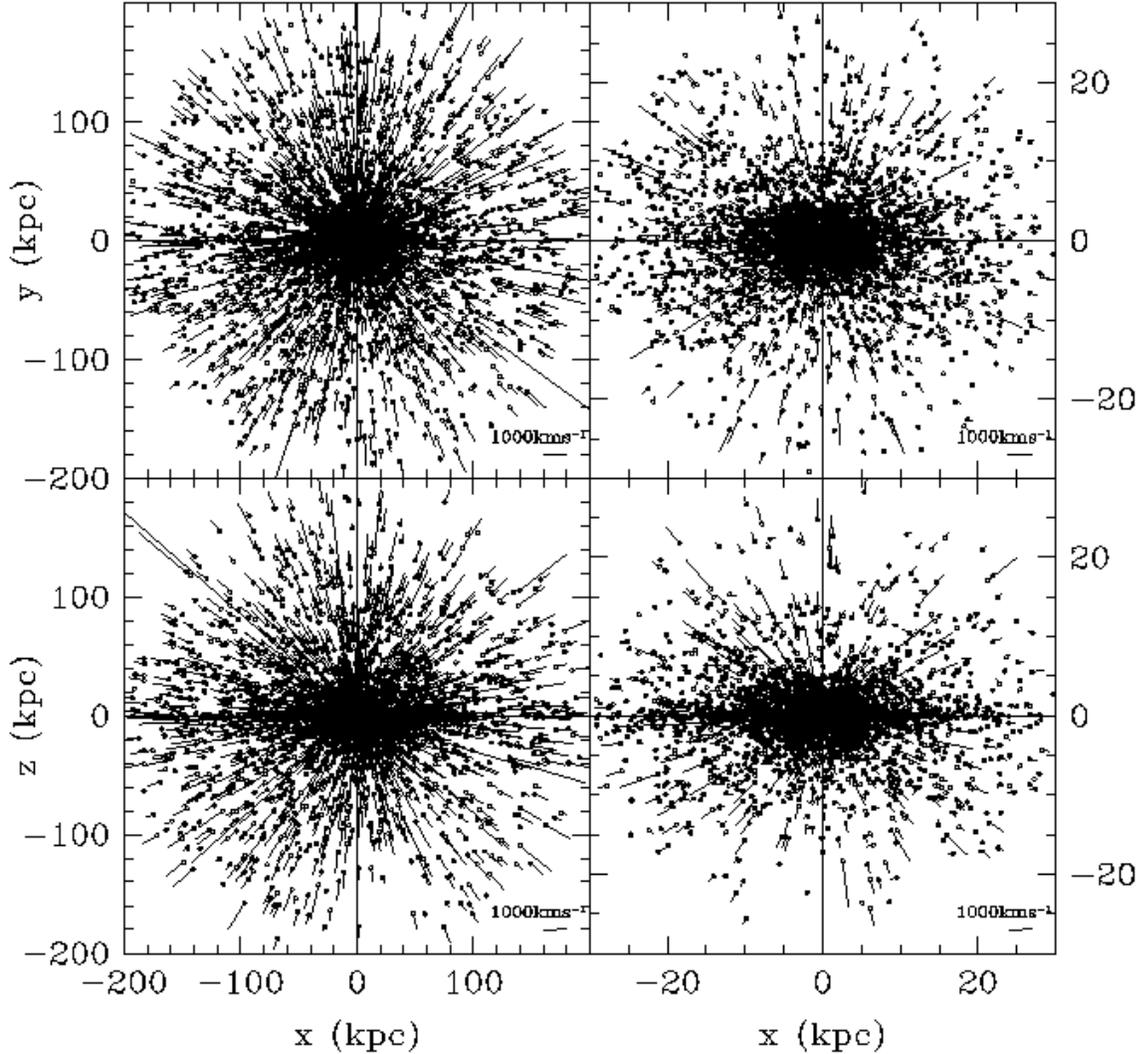}
\caption{Present-day spatial distribution and velocity vectors of HVSs ejected by a binary
BH at the Galactic center. The length of each vector is proportional to speed.
Ten thousand HVSs were generated at a constant rate in the past $10^9\yr$, with a
velocity and spatial distribution obtained from the three-body scattering experiments
of \citet{SHM06}. The binary has a mass $3.6\times 10^6\msun$, mass ratio of 1/81, orbital
semimajor axis equal to $0.1a\h$, and eccentricity 0.3, and orbits in  the
$(x,y)$ plane. The reference axes are set along the $(x,y,z)$-axes of
the triaxial halo potential in eq. (\ref{eq:Phihpq}).
} \label{fig:map} \end{center} \end{figure*}

\begin{figure*}
\begin{center}
%\epsscale{0.8} \plotone{alphar.epsi}
\includegraphics[width=0.9\textwidth,angle=0]{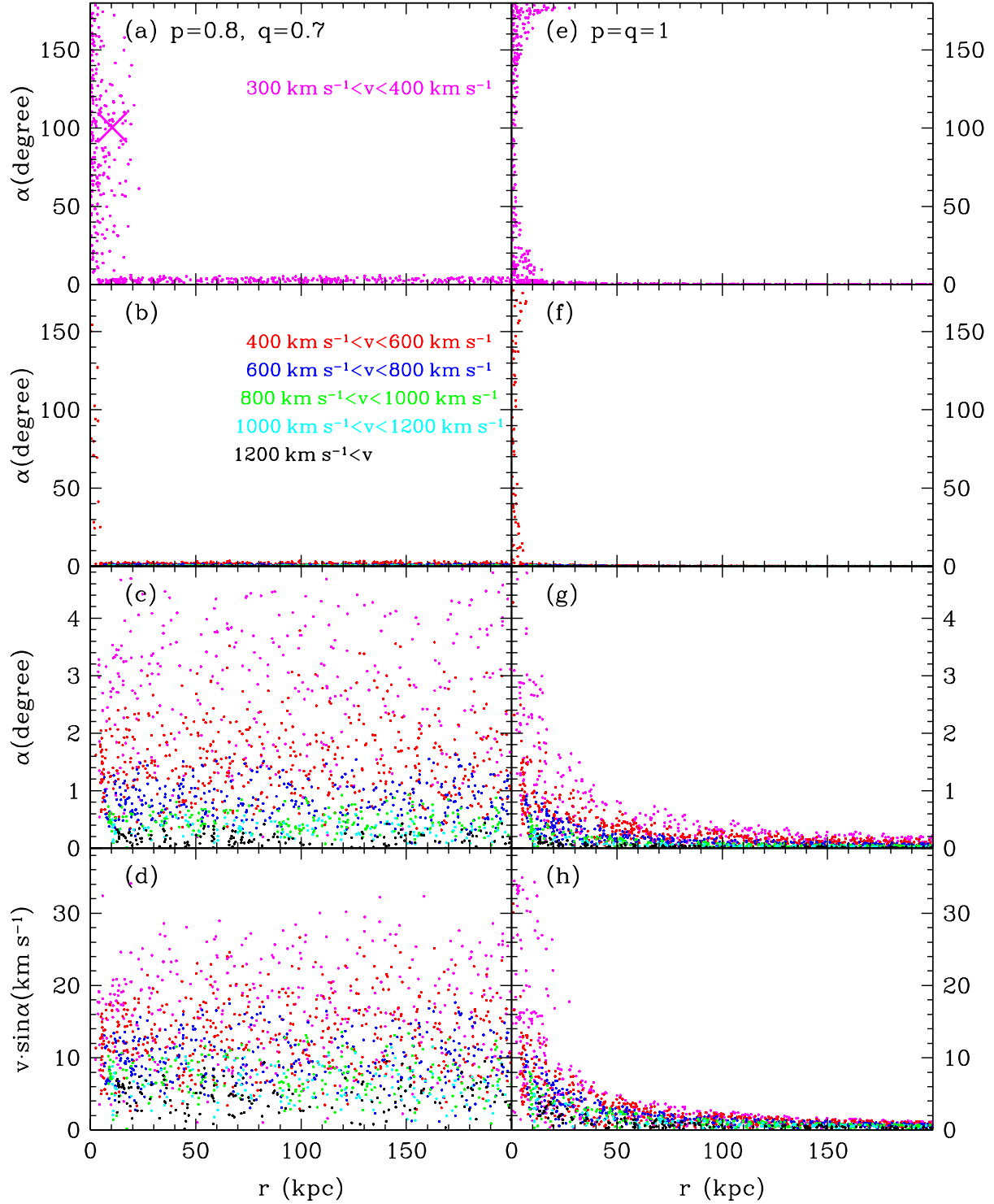}
\caption{Panels (a)-(c): Deflection angle $\alpha=\arcsin(|\vec{r}\times\vec{v}|/rv)$
between velocity and position vectors for all HVSs plotted in Fig. \ref{fig:map},
as a function of Galactocentric distance $r$. The panels  show different velocity
ranges or different scales in $\alpha$.  Panel (d):  transverse velocity in the Galactocentric
frame versus distance $r$ for all stars shown in Panel (c). {\it Panels (a)--(d)}:
triaxial halo potential with $p=0.8$ and $q=0.7$  (see eq. \ref{eq:Phihpq}).
{\it Panels (e)--(h)}: spherical halo potential with $p=q=1$.
Different colors depict different velocity ranges: 300-400$\kms$
({\it magenta}), 400--600$\kms$ ({\it red}), 600--800$\kms$ ({\it blue}), 800--1000$\kms$
({\it green}), 1000--1200$\kms$ ({\it cyan}), $>1200\kms$ ({\it black}).
HVSs with angles between $5^\circ$ and $180^\circ$
in Panels (a), (b), (e) and (f) are bound stars with significantly bent orbits, and their
detailed $\alpha$ distribution depends on halo triaxiality. Stars with small deflections
($\alpha \la5^\circ$) are either unbound (and have distances $\ga 10\kpc$) or
are in the initial phases of their orbital period (and are closer to the Galactic center, see
also Fig. \ref{fig:alphat}). The large cross in Panel (a) flags the locus of
the bound star whose orbit is shown in Fig. \ref{fig:orbit}.
} \label{fig:alphar} \end{center} \end{figure*}

\subsection{Initial conditions}\label{subsec:initial}

According to the study of \citet{YT03}, three-body interactions between 
ambient  stars and a BH pair (where the secondary BH may be an IMBH
inspiralling towards Sgr A$^*$; e.g., \citealt{BGPZ06}) expel HVSs 
($v>10^3\kms$) at a rate that can be as large as $\sim 10^{-4}\yr^{-1}$ 
(for a binary with semimajor axis $0.5\times 10^{-3}\pc$ and mass ratio of 0.01).
Tidal break-up of binary stars (``Hills' mechanism'') ejects 
HVSs at a rate $\sim 10^{-5}(\eta/0.1)\yr^{-1}$,
where $\eta$ is the fraction of stars in binaries with semimajor axis $\la 
0.3\,\au$. Close encounters of two single stars in the vicinity of Sgr A$^*$ may
also produce HVSs but at the negligible rate of $10^{-11}\yr^{-1}$. 
Different ejection mechanisms give origin to different spatial and velocity
distributions of HVSs. While Hills' mechanism predict HVSs to be expelled 
isotropically at an approximately constant rate, in models involving a BH pair 
HVSs are ejected preferentially within the orbital plane of the binary 
in a short burst lasting a few Myr \citep{ZB01,Levin06,SHM06,SHM07}. In the latter
case the degree of anisotropy depends on binary separation, the mass ratio,
and the orbital eccentricity of the BH binary.

For illustrative purposes, we assume in this paper that HVSs  are ejected  from
the Galactic center by the BH binary mechanism.  We use the stellar spatial and 
velocity distributions derived from the scattering experiments of \citet{SHM06} for a binary
with mass ratio $M_2/M_1=1/81$, semimajor axis $a=0.1\,a_h$, and eccentricity $e=0.3$. 
The binary orbit is in the Galactic disk plane, and the velocity of the lighter hole at 
pericenter is directed along $\phi=3\pi/2$. The ``hardening''  radius $a_h$ is defined as 
\citep{Q96}
\begin{eqnarray}
& & a_h \equiv \frac{GM_2}{4\sigma_c^2} \nonumber \\
&\simeq& 0.39\pc\left(\frac{M_2}{M_1}\right)
\left(\frac{M_1}{3.6\times10^6\msun}\right)
\left(\frac{100\kms}{\sigma_c}\right)^2.
\label{eq:ah}
\end{eqnarray}
When $a>a_h$, the binary 
separation decreases both by dynamical friction and three-body interactions with low-angular
momentum stars passing in its immediate vicinity. After the binary becomes ``hard'' ($a<a\h$), 
the bound pair loses orbital energy mainly through three-body interactions until gravitational radiation takes over \citep{BBR80,Y02}. 
The ejection speed of the stars at infinity spans a range with r.m.s. $\sim
7\times10^2\kms(M_2/0.01M\bh)^{1/2}(10^{-3}\pc/a)^{1/2}(M\bh/3.6\times10^6\msun)^{1/2}$ (see eq.~33 in \citealt{YT03}). In the calculations below, we assume $10^4$
HVSs are ejected from the Galactic center at a constant rate over the last  $10^9$ years, 
and ignore for simplicity the orbital evolution of the binary during such timescale.
Different stars move independently in the Galactic potential. A
fourth-order Runge-Kutta method with adaptive stepsize control is used
to solve the differential equations of the motion of the stars.
Note that the
adopted  mass ratio and semimajor axis are within the allowed parameter space for a
BH pair at the Galactic center (see Fig. 2 in \citealt{YT03}).

\subsection{Results} \label{subsec:results}

\begin{figure}
\begin{center}
%\epsscale{0.8} \plotone{orbit.epsi}
%\includegraphics[width=0.8\textwidth,angle=0]{orbit.epsi}
\includegraphics[width=\hsize,angle=0]{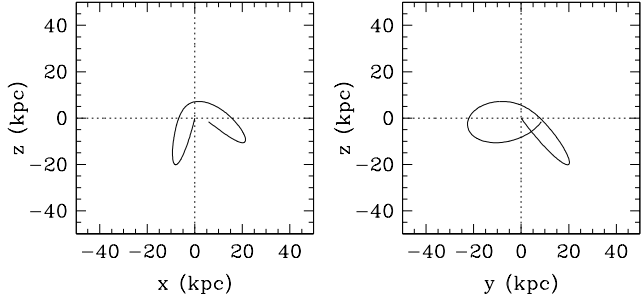}
\caption{The bound orbit of a simulated HVS ejected from the Galactic center. The 
orbit is significantly bent, rather than radial as expected in a spherical potential.
The star's travel time $820\Myr$, and its present-day velocity and Galactocentric distance
are $300\kms$ and $10\kpc$, respectively.
}
\label{fig:orbit} \end{center} \end{figure}

\begin{figure}
\begin{center}
%\epsscale{0.6} \plotone{histogram.epsi}
%\includegraphics[width=0.6\textwidth,angle=0]{histogram.epsi}
\includegraphics[width=\hsize,angle=0]{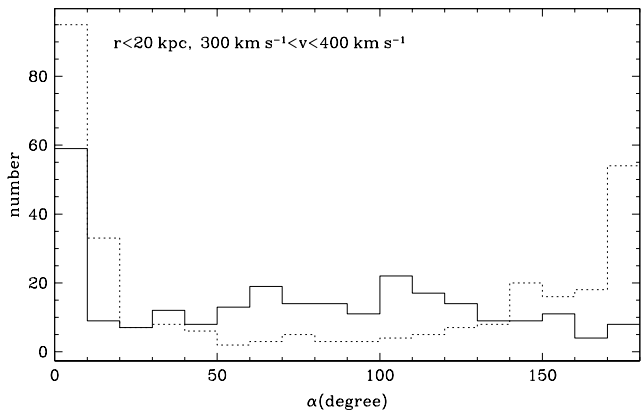}
\caption{Histogram of the deflection angle distribution of all stars shown in
Fig.\ \ref{fig:alphar}(a) ({\it solid line}) and (e) ({\it dotted line}) with
$r<20\kpc$ and $300\kms<v<400\kms$. The bound stars are clustered around
$0^\circ$ and $180^\circ$ in a spherical halo potential ({\it dotted line}),
while their distribution is spread uniformly over the range $10^\circ$--$180^\circ$
in a triaxial halo potential ({\it solid line}).}
\label{fig:histogram} \end{center} \end{figure}

\begin{figure}
\begin{center}
%\epsscale{0.6} \plotone{alphat.epsi}
%\includegraphics[width=0.6\textwidth,angle=0]{alphat.epsi}
\includegraphics[width=\hsize,angle=0]{alphat.epsi}
\caption{Deflection angle versus travel time from the Galactic center. 
{\it Top panel:} all stars shown in Fig. \ref{fig:alphar}(a) with $r<50\kpc$ and
$300\kms<v<400\kms$. {\it Bottom panel:} all stars shown in Fig. 
\ref{fig:alphar}(b) with $r<200\kpc$ and $400\kms<v<600\kms$. HVSs
with short travel times, $\la 50$ Myr ({\it top}) and $\la 400\Myr$ ({\it bottom})
have $\alpha$ angles smaller than $5^\circ$.  The large cross in the top panel
flags the locus of the bound star whose orbit is
shown in Fig. \ref{fig:orbit}.
}
\label{fig:alphat} \end{center} \end{figure}

We use the initial conditions described above and numerically integrate the orbits of 
HVSs in the Galactic potential. Figures \ref{fig:map}  and \ref{fig:alphar} show  maps of 
stellar position and velocity vectors at the present time and their deflection angles
$\alpha=\arcsin(|\vec{r}\times\vec{v}|/rv)$.  For HVSs with $v\ga 600\,\kms$, deviations
are quite small, $\alpha\la 5^\circ$, at all distances within $200\,\kpc$.  Lower velocity 
stars at small distances can instead be bound, and their deflection angles extend to 
$180^\circ$. Many of the stars with $300\kms\la v\la 400\kms$ and $r\la 20\,\kpc$ in 
Figure \ref{fig:alphar}(a) follow bound trajectories, 
while  no stars in this velocity range have substantial deviation angles at large 
Galactocentric distances.  One example of a bound orbit is shown in
Figure \ref{fig:orbit}, where the trajectory has been significantly bent by the
triaxial halo and the flattened disk potentials, and the star does not return to the Galactic center.  
Note that bound stars with $v>400\,\kms$ and large $\alpha$'s are typically 
observed at smaller distances ($r\la 10\,\kpc$) than bound stars of lower velocities.
This is because a larger ejection speed from the Galactic center is needed to 
maintain a high velocity at large distances, and such stars may then  have either escaped
from the halo or not have had sufficient time to come back and show a significant
bend in their orbits. (For reference, the local escape speed is about 
$500-600\,\kms$, \citealt{Smith06}.) We find that 60\% of the stars with $v>300\kms$ at 
$r<20\kpc$ have velocities $v<400\kms$, and 20\% of the stars with $v>300\kms$ at
$r<10\kpc$ have velocities $v<600\kms$.  

\begin{figure*}
\begin{center}
%\epsscale{0.8} \plotone{mapJn.epsi}
\includegraphics[width=0.9\textwidth,angle=0]{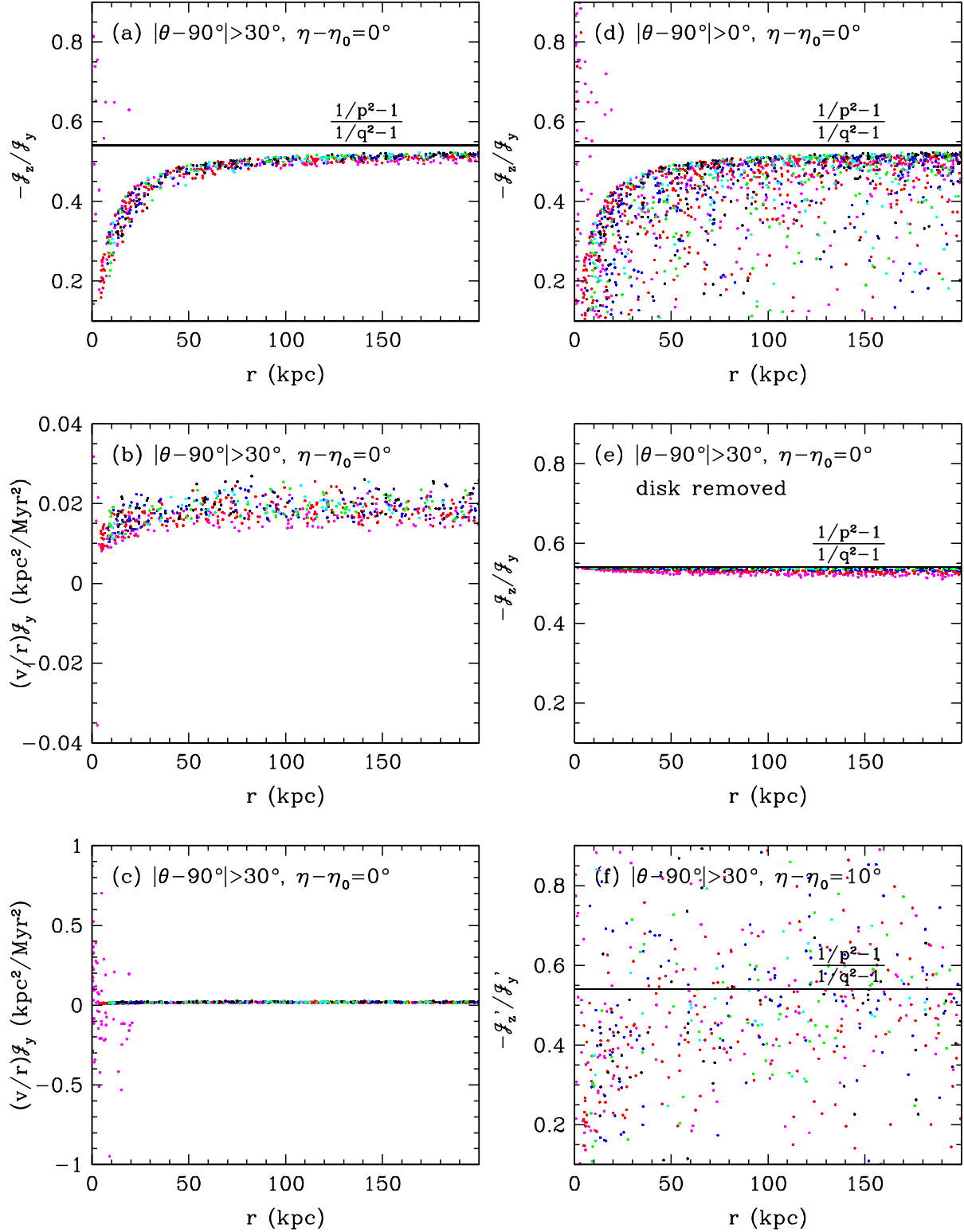}
\caption{The variables $-\calJ_z/\calJ_y$ and $(v/r)\calJ_y$  versus
Galactocentric distances $r$ for all HVSs far from the plane of the
disk, $|\theta-90^\circ|<30^\circ$. Stars near the plane are
included only in Panel (d). Different colors represent different velocity
ranges as in Fig. \ref{fig:alphar}. Panel (c) shows $\calJ_y$ on a different
scale than Panel (b), and the scattered magenta
dots are bound stars with significantly bent orbits. Panel (e) shows the
effect of removing the disk potential from the calculation. Panel (f) shows
the results in a reference frame with $\eta-\eta_0=10^\circ$, different from
the triaxial frame of the halo potential.  The solid line represents the
triaxiality of the halo potential $T=0.54$ (with $p=0.8$ and $q=0.7$ in
eq.~\ref{eq:Tdef}) assumed in the calculation.
} \label{fig:mapJn} \end{center} \end{figure*}

Figure \ref{fig:alphat} shows that bound stars with large deflection angles 
($5^\circ\la\alpha\la180^\circ)$ have generally traveled a long time after 
ejection ($\ga$50 Myr for $r\la 20\,\kpc$), and many of them have experienced at least one
orbital period (Fig. \ref{fig:orbit}). HVSs with small deflections ($\alpha\la5^\circ$) 
have instead a short travel time and are generally on the initial phases of their 
first orbital periods (see the concentration of the stars at the left bottom of the 
panels in Figure \ref{fig:alphat}). The transverse velocities (in the Galactocentric
frame) of stars with $v>300\kms$ are typically higher than $3\,\kms$, and can be up to 
$30\,\kms$ (hundreds of $\kms$) for unbound (bound) HVSs. Note that
$3\,\kms$ corresponds at a distance of 100 kpc to a proper motion of $20\,\mu$as in 
three years, which can be resolved by the next generation of astrometric
surveys like GAIA. According to our calculations most HVSs would have transverse 
velocities $\ga 3\,\kms$ even in the case of a weakly triaxial halo with
$(p,q)=(0.95,0.9)$: these transverse velocities are larger than those
associated with the bending of stellar trajectories caused by the axisymmetric disk
(see Fig.\ \ref{fig:alphar}h).

A comparison between Figures \ref{fig:alphar}(e)--(g) and \ref{fig:alphar}(a)--(c)
shows that the distribution of deflection angles versus distance is different 
for trajectories in a spherical rather than triaxial halo. In 
Figure \ref{fig:alphar}(e) and (f), the $\alpha$ angles of bound stars are 
clustered around $0^\circ$ and $180^\circ$ because their orbits are
highly eccentric with little bending (see also Fig.~\ref{fig:histogram}). By contrast, in Figure \ref{fig:alphar}(a) and
(b) the deflection angles of bound stars lies at intermediate values.
This difference can be used as an indicator of the triaxiality of the MW
dark matter halo.

\begin{figure*}
\begin{center}
%\epsscale{0.8} \plotone{mapJns.epsi}
\includegraphics[width=0.8\textwidth,angle=0]{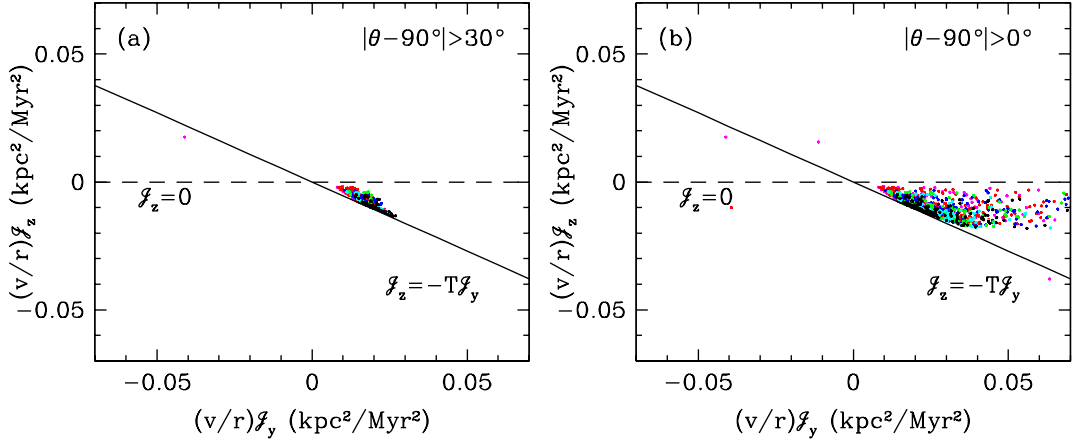}
\caption{The variable $(v/r)\calJ_z$ versus $(v/r)\calJ_y$. The solid and
dashed lines correspond to $\calJ_z=-T\calJ_y$ and $\calJ_z=0$ (see
eq.~\ref{eq:calJaxisy}), respectively.  The stars are those shown in Fig.
\ref{fig:mapJn}.  Panel (a) includes only stars far from the plane of the
disk, $|\theta-90^\circ|>30^\circ$, while Panel (b) includes all stars.
The effect of the disk potential can be gauged from the spread of the stellar
dots from the solid line to the dashed line in Panel (b). The magenta and red dots
scattered above the dashed line or below the solid line represent bound stars with
bent orbits.
} \label{fig:mapJns} \end{center} \end{figure*}

Figure \ref{fig:mapJn}(a)--(c) depicts the values of $-\calJ_z/\calJ_y$ and
$(v/r)\calJ_y$ versus distance of all the HVSs plotted in Figure \ref{fig:map}
having $v>300\kms$ and far from the plane of the disk, i.e. with $|\theta-90^\circ|>30^\circ$.
Figure \ref{fig:mapJn}(d) shows all stars with $\theta$ in the range
$0^\circ$--$180^\circ$. As seen in Figure \ref{fig:mapJn}(a), the ratio
$-\calJ_z/\calJ_y$ at large distances ($r\ga 100\kpc$) is close to the
triaxiality of the halo potential ($T=0.54$, {\it solid line}) assumed in our
calculations, with only a small scatter. The scatter is larger in the quantity
$(v/r)\calJ_y$ plotted in Figure \ref{fig:mapJn}(b). The slight offset ($\la10\%$)
of the stellar dots from the solid line is partly due to the disk
potential that causes an additional bending of stellar trajectories towards the disk
plane, increasing $|\calJ_y|$ without changing $\calJ_z$ (see the Appendix for details about
the correction of such offset owing to the Galactic disk). As shown in Figure \ref{fig:mapJn}(e), 
HVSs become better tracers of triaxiality after removing the disk potential from our
calculations. The higher the velocities, the smaller the offset. The curvature of the
ratio $-\calJ_z/\calJ_y$ towards smaller values at small distances is also
an effect of the disk. The ratio $-\calJ_z/\calJ_y$ for stars close to disk plane is not a good
approximation of halo triaxiality even at large distances (see Fig.\ \ref{fig:mapJn}d).
We plot the values of $\calJ_z$ versus $\calJ_y$ in Figure \ref{fig:mapJns}, where Panel
(a) shows stars with $|\theta-90^\circ|<30^\circ$, and Panel (b) stars with $\theta$ in
the full range $0^\circ$--$180^\circ$. The effect of the disk potential can be seen
in the spreading of the stellar dots from the solid line $\calJ_z=-T\calJ_y$ to the
dashed line $\calJ_z=0$.
Note that the scatter in the ratio $-\calJ_z'/\calJ_y'$ increases
significantly if the reference frame is different from the frame of the triaxial halo
potential (Fig. \ref{fig:mapJn}f).

\section{Summary}\label{sec:conclusion}

We have studied the unique kinematics of HVSs ejected from the 
Galactic center with almost zero initial specific angular momentum.
HVSs can travel in the Galactic halo on either bound or unbound orbits,
and their spatial and velocity distribution 
at large Galactocentric distances ($r\ga50\kpc$) contain information
on the asphericity of the halo gravitational potential. We have
proposed an estimator of the triaxiality of the Galactic dark matter 
halo that is determined solely by instantaneous position and velocity 
vectors of HVSs, is independent of the details of the ejection mechanism, 
and does not require an accurate knowledge of halo mass.
Future astrometric and deep wide-field surveys of HVSs should detect significant numbers
of HVSs, which could be used to determine the triaxiality of the
MW halo by applying the method proposed in this paper.

The new class of possibly bound HVSs with velocities $+275<v_{\rm rf}<+450\,\kms$
recently observed by \citet{Brown07} has Galactocentric distances in the range
30-60$\kpc$ or 10-20$\kpc$ depending on whether they are main-sequence or blue
horizontal branch stars. In the first case (main-sequence stars at large distances),
they have a lifetime of $\la 100\Myr$ and are, according to Figure \ref{fig:alphat},
on the initial phases of their first orbital periods. Their deflection
angles are expected to be rather small, supporting the fact that a significant
excess of B-type stars is observed only at large positive velocities \citep{Brown07}.
If bound HVSs are blue horizontal branch stars instead at smaller distances,
the travel time of the stars can be much longer than $50\Myr$ as at
the ejection moment the stars may not necessarily be blue horizontal
branch stars but at some pre-blue-horizontal-branch stage.
Such stars may have experienced at least one orbital period, and many of them should
be returning to the Galactic center or their orbits should have been significantly
bent by the asymmetric Galactic potential (Fig. \ref{fig:alphat}). This scenario
appear unlikely since it does not agree with the observed positive radial velocities.

\begin{figure}
\begin{center}
%\epsscale{0.8} \plotone{Texample.epsi}
%\includegraphics[width=0.8\textwidth,angle=0]{Texample.epsi}
\includegraphics[width=\hsize,angle=0]{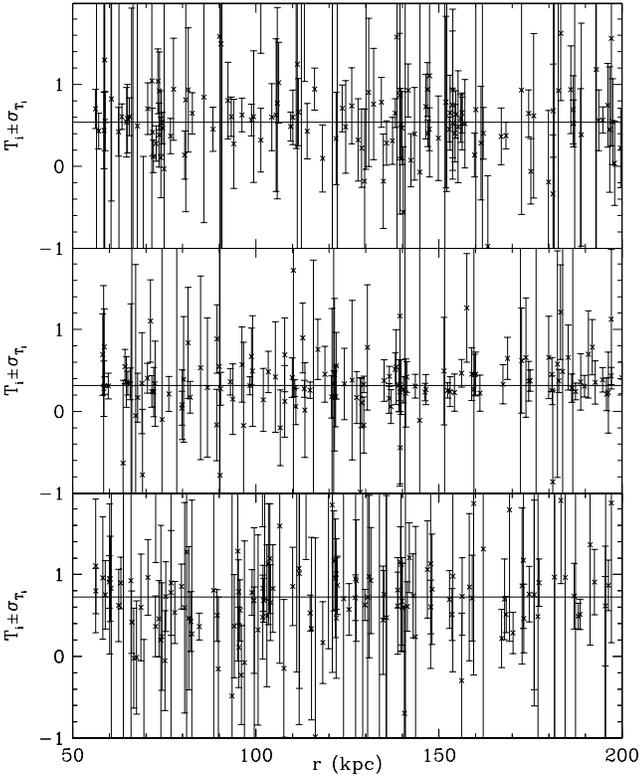}
\caption{Simulated triaxiality parameters and their errors from a mock
sample of HVSs with $v>300\kms$, $|\theta-90^\circ|>30^\circ$, and $55\kpc<r<200\kpc$.
An observational error of $3\kms$ is assumed for in ``measured'' one-dimensional velocity.
From top to bottom, the values of $(p,q)$ used in the calculations
are (0.8, 0.7), (0.8, 0.6), and (0.8, 0.75), corresponding to a triaxiality
$T$=0.54, 0.32, and 0.72, respectively ({\it horizontal lines}).
A statistical analysis of the sample using eqs.
(\ref{eq:Tbar}) and (\ref{eq:Terror}) yields $(\bar{T},\sigma_{\bar{T}})$=(0.50,
0.02), (0.31, 0.01), and (0.64, 0.02), respectively. For clarity, only 1/3
of the sample points are drawn in the figure.}
\label{fig:Texample} \end{center} \end{figure}

It is interesting at this stage to provide an example of a statistical estimate of 
halo triaxiality from a mock sample. Let us assign a measurement error of 
$\sigma_a=3\,\kms$ ($a=x,y,z$) 
to the one-dimensional velocities of all stars in Figure \ref{fig:mapJn}(a) 
having $v>300\kms$, $|\theta-90^\circ|>30^\circ$, and $55\,\kpc<r<200\,\kpc$
(our error analysis assumes that the distances to HVSs are known to within
ten percent).
We have simulated the observed velocities of such a sample, and plot in 
the top panel of Figure \ref{fig:Texample} the values of 
$T_i$ and $\sigma_{T_{i}}$ derived for each HVS. Using equations (\ref{eq:Tbar}) and
(\ref{eq:Terror}), we obtain $\bar{T}=0.50$ and $\sigma_{\bar{T}}=0.02$. 
For $\sigma_a=5\,\kms$, we obtain $\bar{T}=0.50$ and $\sigma_{\bar{T}}=0.03$.
These errors are comparable with the systematic error caused by the 
flattened disk (see the slight offset of dots from the solid line in Fig. 
\ref{fig:mapJn}a). We have tried different values of the gravitational potential 
parameters $(p,q)$ in the calculations, and found that the estimated value of $\bar{T}$ 
is always consistent with the assumed halo triaxiality (see middle and bottom panels in Fig.
\ref{fig:Texample}).

\begin{figure}
\begin{center}
%\epsscale{0.8} \plotone{variance.epsi}
%\includegraphics[width=0.8\textwidth,angle=0]{variance.epsi}
\includegraphics[width=\hsize,angle=0]{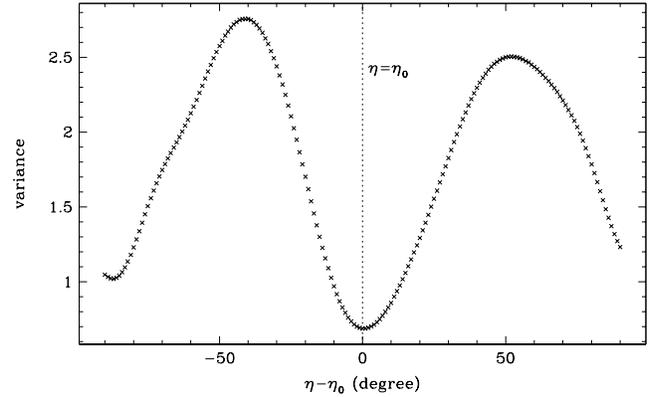}
\caption{Variance of the triaxiality parameter $T'$ of the simulated sample (top panel
in Fig. \ref{fig:Texample}) in different reference frames as a function of
$\eta-\eta_0$. Here $\eta_0$ is the angle measured
counter-clockwise from a reference direction (e.g. the line from the Galactic
center to the Sun) to the $x$-axis of the halo potential, and
$\eta$ is the angle formed by the observational frame with the reference direction.
The variance has a minimum at $\eta-\eta_0=0^\circ$.} \label{fig:variance} \end{center} \end{figure}

If the $x$-axis of the halo potential form an angle $\eta_0$ (measured counter-clockwise)
with the reference axis of the observations, then (as seen from Fig. \ref{fig:variance}) 
the minimum variance of $\calJ_{z,i}'/\calJ_{y,i}'$ in a set of simulated samples
each in a frame at angle $\eta$ from the observational reference axis 
(see eq.~\ref{eq:variance}) occurs for $\eta=\eta_0$. Our calculations show that 
an error of $5^\circ$ in the estimate of $\eta_0$ may cause an error of 
2\% in the estimate of $\bar{T}$.

Note that the axis ratios $(p,q)$ of the halo potential are
degenerate in the defined triaxiality parameter (eq.~\ref{eq:Tdef}). After 
determining $T$, the values of $(p,q)$ could be also obtained by using any
value of $(\calJ_x, \calJ_y,\calJ_z)$ (e.g., see Fig.~\ref{fig:mapJn}b), but
the modeling would be sensitive to halo mass and the shape of halo potential
used (see also the determination of axis ratios in \citealt{Gnedin05}
by tracing back HVS orbits).

Finally, it is possible that a few HVSs in the halo may be 
produced by the interactions of stars with an IMBH in satellite galaxies like
the Large Magellanic Cloud \citep{Edelmann05,GPZ07}, and that these 
would contaminate the sample ejected from the Galactic center. The ejection
rates from the satellite dwarfs of the MW are expected to be much smaller than the rate from 
Sgr A$^*$, however. Such ``satellite'' HVSs will also have much larger angular 
momenta in the Galactocentric rest-frame, and should be easily distinguishable 
from the Galactic center sample.

%\acknowledgements

We have benefited from discussions with Francesco Haardt, Youjun Lu, and Scott 
Tremaine. We thank Alberto Sesana for providing the initial velocity and spatial 
distributions of HVSs ejected from a black hole binary. P.M. acknowledges 
financial support from NASA through grants NAG5-11513 and NNG04GK85G, and 
from the Alexander von Humboldt Foundation. Q.Y. acknowledges initial support 
from NASA through Hubble Fellowship grant HST-HF-01169.01-A awarded by the Space Telescope
Science Institute, which is operated by the Association of Universities for
Research in Astronomy, Inc., for NASA, under contract NAS 5-26555.

\appendix

\section{Correction of the effect of disk potential on the inferred
halo triaxiality}

As shown in Figure \ref{fig:mapJn}(a), the triaxiality obtained through
$T=-\calJ_z/\calJ_y$ has a small offset from the true halo triaxiality. Part of
this offset comes from the effects of the flattened disk potential, and part from
the fixing of $(\theta,\phi)$ in the integration of equation (\ref{eq:TdJ}).
The offset due to the constant $(\theta,\phi)$ assumption is smaller for stars
of higher velocities, as seen in Figure \ref{fig:mapJn}(e) that shows the
results after the disk potential was removed.  Below we provide a method to correct
the offset due to the Galactic disk. As in the determination of halo triaxiality 
proposed in this work, even for this correction it is not necessary to trace back the 
orbits of HVSs.

The contribution to the change in specific angular momentum due to the disk
potential can be expressed as:
\be
J_{y,}{\disk}=\int^{r_0}_{r_{\rm ej}}\frac{dJ_{y,}{\disk}}{dt}\frac{dt}{dr}dr,
\ee
where
\be
\frac{dJ_{y,}{\disk}}{dt}=2xz\left(\frac{\partial\Phi\disk}{\partial z^2}-\frac{\partial\Phi\disk}{\partial x^2}\right),
\label{eq:dJdisk}
\ee
\be
\frac{dt}{dr}=\frac{1}{v_r}\simeq \frac{1}{\sqrt{v^2_0+2{\Phi}(\vec{r}_0)-2{\Phi}(\vec{r})}},
\label{eq:vr}
\ee
$r_{\rm ej}$ is the initial distance of the HVS from the Galactic center at
ejection, $\vec{r}_0$ and $v_0$ are the current position and velocity of the
HVS, $r_0=|\vec{r}_0|$, and $v_r$ is its radial velocity at $r$.
%In equation (\ref{eq:vr}), the radial velocity component dominates. 
We can remove the effect of the disk potential from the variable $\calJ_y$ by
computing $\calJ_y-\calJ_{y,}{\disk}$, where
\begin{eqnarray}
& & \calJ_{y,}{\disk}  =  \frac{J_{y,}{\disk}}{\sin\theta\cos\phi\cos\theta}\nonumber \\
& = & \int^{r_0}_{r_{\rm ej}}\frac{G M\disk}{(r^2+a\disk^2+b\disk^2+2a\disk
\sqrt{r^2\cos^2\theta+b\disk^2})^{3/2}} \nonumber\\ 
& & \frac{a\disk r^2}
{\sqrt{r^2\cos^2\theta+b\disk^2}}\frac{dr}{v_r},
\label{eq:calJydisk}
\end{eqnarray}
and where equation (\ref{eq:Phid}) describing the disk
potential has been used. The angles $(\theta,\phi)$ are fixed in the integration
to be present-day values. In addition to the disk potential, we also need to assume a
halo potential to determine $v_r$, for which we may use the spherical part of
the halo potential by setting $(p,q)=(1,1)$.
We have tested this correction and found that half of the offset to the true
halo triaxiality can be corrected.

\end{document}